# How to develop, externally validate, and update multinomial prediction models

Celina K. Gehringer*[1,2], Glen P. Martin[3], Ben Van Calster[4,5], Kimme L. Hyrich[1,6], Suzanne M. M. Verstappen[1,6], Jamie C. Sergeant[1,2]

[1]Centre for Epidemiology Versus Arthritis, Centre for Musculoskeletal Research, Division of Musculoskeletal and Dermatological Sciences, University of Manchester, Manchester, UK

[2]Centre for Biostatistics, Manchester Academic Health Science Centre, University of Manchester, Manchester, UK

[3]Centre for Health Informatics, Division of Informatics, Imaging and Data Sciences, University of Manchester, Manchester, UK

[4]Department of Biomedical Data Sciences, Leiden University Medical Centre, Leiden, the Netherlands

[5]Department of Development & Regeneration, KU Leuven, Leuven, Belgium

[6]NIHR Manchester Biomedical Research Centre, Manchester University NHS Foundation Trust, Manchester Academic Health Science Centre, Manchester, UK

*Corresponding author: Celina K. Gehringer
Centre for Epidemiology Versus Arthritis
Stopford Building, Oxford Road
M13 9PG, Manchester
University of Manchester
United Kingdom
Email: celina.gehringer@postgrad.manchester.ac.uk

Funding: This work was funded by Versus Arthritis (grant number 21755). KLH is supported by the NIHR Manchester Biomedical Research Centre.

Conflict of interest: KLH has received speaker honoraria from Abbvie and grant income from BMS and Pfizer. The remaining authors have no conflicts of interest to declare.

**Abstract**

**Objectives:** Multinomial prediction models (MPMs) have a range of potential applications across healthcare where the primary outcome of interest has multiple nominal or ordinal categories. However, the application of MPMs is scarce, which may be due to the added methodological complexities that they bring. This article provides a guide of how to develop, externally validate, and update MPMs.

**Study Design and Setting:** Using a previously developed and validated MPM for treatment outcomes in rheumatoid arthritis as an example, we outline guidance and recommendations for producing a clinical prediction model using multinomial logistic regression. This article is intended to supplement existing general guidance on prediction model research.

**Results:** This guide is split into three parts: 1) Outcome definition and variable selection, 2) Model development, and 3) Model evaluation (including performance assessment, internal and external validation, and model recalibration). We outline how to evaluate and interpret the predictive performance of MPMs. R code is provided.

**Conclusions:** We recommend the application of MPMs in clinical settings where the prediction of a nominal polytomous outcome is of interest. Future methodological research could focus on MPM-specific considerations for variable selection and sample size criteria for external validation.

## 1. Background

Clinical prediction models (CPMs) use patient characteristics to estimate an individual's risk of having (diagnostic models) or developing (prognostic models) an outcome of interest at a particular time point(1). Across healthcare, there are many instances where the outcome of interest is polytomous, meaning that it has more than two mutually exclusive categories, which can either be ordinal (ordered) or nominal (unordered). For example, it is of interest to diagnose ovarian tumours as benign, borderline, invasive, or metastatic, rather than as simply benign vs malignant(2), which acknowledges the heterogeneity between malignant tumours(3,4). In the machine learning literature, such outcomes are referred to as multi-class or multi-category(5–11). There are a range of methods that could be used to model polytomous outcomes(12), with multinomial logistic regression (MLR) being a common statistical approach for developing CPMs(12–14). Multinomial prediction models (MPMs) are any risk prediction models for polytomous outcomes, and this article focuses exclusively on guidance for MPMs for a nominal outcome developed using MLR. An overview of alternative methods for modelling polytomous outcomes is provided in articles by Edlinger *et al.*(12)(ordinal outcomes), Pate *et al.*(15)(nominal outcomes), van Calster *et al.*(9) and Kruppa *et al.*(16,17)(machine learning).

MPMs have been applied to several clinical settings(2,18–21), however, the adoption of this approach is still relatively scarce(13). We hypothesise that a reason for the lack of application of MPMs is lack of awareness of these methods, and the added complexities they bring, particularly when evaluating performance. Indeed, there has been a range of emerging methodological research around MPMs, including sample size considerations(22,23), the assessment of discrimination(24) and calibration(25), and guidance on validation and updating(26). However, it can be challenging to navigate this methodological literature. We are aware of one paper that provides guidance regarding measures for discrimination performance(27). Therefore, the aim of this article is to provide more extended and up-to-date guidance for statisticians, epidemiologists, and their collaborators on how to develop, externally validate, and update MPMs.

## 2. Clinical example

In a previous publication(28), we reported the development and validation of a MPM for methotrexate (MTX) treatment outcomes in rheumatoid arthritis (RA). We now use this example to illustrate concepts in the current paper. RA is a heterogenous autoimmune condition that causes swelling and stiffness in the joints of the hands and feet(29). Pharmacological management of RA should be commenced as soon as diagnosis is made, with MTX being the recommended first-line therapy(30,31). However, response to MTX is not universal, with one study reporting that 43% of patients do not respond to treatment by 6 months(32). Discontinuation of treatment due to adverse events (AEs, i.e., gastrointestinal events(33)) is also common, occurring in 20-30% of patients in the first year of taking MTX(34–36). Identification of patients that are at high risk of non-response or discontinuation due to AEs is therefore of clinical importance so that disease control can be expedited using alternative treatments. We previously developed a MPM(28) to estimate an individual's risk of 1) not achieving the defined state of low disease activity (LDA)(37) at 6 months, or 2) achieving LDA at 6 months, or 3) discontinuing due to AEs within 6 months of commencing MTX. This complementary paper aims to provide a broader view of translating the methodological advances surrounding MPMs into practical guidance for applied researchers. To help maximise the impact of this guidance, we encourage readers to also consult the previous paper(28) with details of the methodological approaches used in the RA case study.

## 3. How to develop, validate, and update a multinomial prediction model

This section will provide a guide of how to develop, validate, and update MPMs, with computational guidance provided in Section 3.4. Note that the development and validation of any CPM, multinomial or otherwise, should follow best practice guidelines(38) and be reported using the TRIPOD (Transparent Reporting for a multivariable prediction model for Individual Prognosis Or Diagnosis)

checklist(38,39). In this article, we focus solely on the additional considerations for MPMs. All analyses were carried out in R(40) (version 4.1.2) and code for the clinical example is provided on GitHub(41).

## 3.1. Outcome definition and variable selection

### *3.1.1. Ordinal vs Nominal*

Ordinal polytomous outcomes have ordered categories, for example the extent of coronary artery disease, where the diagnosis takes any of five increasingly severe categories(42). These outcome types can be modelled using methods that assume proportional odds, which require fewer parameters than MLR models and are therefore at lower risk of overfitting, but at higher risk of severe miscalibration(12). Using MLR to model ordinal outcomes has been suggested as an alternative (ignoring ordinality)(12), which requires larger samples to obtain reliable risk predictions(22,23) (more in Section 3.2.2). The remainder of this article will focus on models developed using MLR and outcomes treated as nominal polytomous.

### *3.1.2. Outcome definition*

Prior to model development, the outcome of a MPM should be carefully defined. A key consideration is whether different outcome categories are linked to different management options, in which case they should be kept separate. A diagnostic example is the assessment of ovarian tumours as benign, borderline, invasive, or metastatic(2), which may be managed differently. In a prognostic setting, a multinomial outcome could also incorporate potential competing risks to the primary outcome, if survival competing risks methods(43,44) are not feasible or necessary. This was the case in our RA example, where continuous-time competing risks approaches were inappropriate in what was essentially a discrete-time setting.

In a MPM, a reference outcome category will need to be defined and the choice of this reference category depends on the clinical area. Often it is the best health outcome from the patient's perspective. The choice of reference category is important for the interpretation of the regression coefficients, but does not affect the predicted probabilities of the model(45). We recommend that patients and clinicians are included in the process to define the reference category.

#### *3.1.3. Candidate predictor variables*

Candidate predictor variables (i.e., those for consideration in the model prior to any data-driven variable selection) can be chosen based on predictive relationships reported in the literature, clinical expertise, availability in clinical care, measurement heterogeneity(46), and causality (especially for prognostic outcomes)(47).The number of candidate predictors included during model development should be informed by the available sample size, relative to the minimum required sample size(22)(more in Section 3.2.2). It may be the case that developing a model is not feasible given the sample size, as it does not allow inclusion of known important predictors. To produce a parsimonious model, data-driven methods can be implemented, for which previously outlined principles(48,49) would also apply for MPMs. These include selecting variables based on significance level (i.e., p-value), information criteria (i.e., AIC, BIC), and penalised likelihood (i.e., LASSO)(more in Section 3.2.4).

Selecting variables for a MPM in a data-driven way can be considered for the model overall or, if the outcome categories are known to have different predictors, separately for each of the dichotomous submodels (or outcome pair, Section 3.2.1)(21). The final set of variables for the MPM would then consist of the variables that were selected in any of the submodels(21). We note that, in general, data-driven variable selection should be avoided where possible, and that it is currently difficult to provide specific guidance as variable selection for MPMs is relatively under-researched.

### 3.2. Model development

#### *3.2.1. Multinomial prediction model equation*

Unlike a binary logistic regression model, there are multiple submodels (or outcome pairs) within a MPM developed using MLR. For an outcome with $k$ categories, the MPM has $k$-1 submodels, which compare each outcome category (Y=2, Y=3…Y=$k$) to the chosen reference category (typically Y=1).

Box 1 presents the general equation for calculating the linear predictors (LPs) for a three-category outcome, as well as applying this to the RA example, where five predictor variables were included.

Each submodel has its own LP, with a model intercept $\beta_{0,i}$ , predictor variables $x_1 \ldots x_p$ and corresponding regression coefficients $\beta_{1 \ldots p}$:

$$LP_1 = log\left(\frac{P(Y=2)}{P(Y=1)}\right) = \beta_{0,1} + \beta_{1,1}x_1 + \beta_{2,1}x_2 + \cdots \beta_{p,1}x_p$$

$$LP_2 = log\left(\frac{P(Y=3)}{P(Y=1)}\right) = \beta_{0,2} + \beta_{1,2}x_1 + \beta_{2,2}x_2 + \cdots \beta_{p,2}x_p$$

The LPs can then be used to calculate predicted probabilities for each outcome category:

$$P(Y=1) = \frac{1}{1 + exp\,(LP_1) + exp\,(LP_2)}$$

$$P(Y=2) = \frac{exp\,(LP_1)}{1 + exp\,(LP_1) + exp\,(LP_2)}$$

$$P(Y=3) = \frac{exp\,(LP_2)}{1 + exp\,(LP_1) + exp\,(LP_2)}$$

In our clinical example, where no LDA is the reference category, the model is written as:

$$LP_1 = log\left(\frac{P(LDA)}{P(no\ LDA)}\right) = \beta_{0,1} + \beta_{1,1}Age + \beta_{2,1}DAS28 + \beta_{3,1}RF\ +\ \beta_{4,1}sex +\ \beta_{5,1}HAQ$$

$$LP_2 = log\left(\frac{P(AEs)}{P(no\ LDA)}\right) = \beta_{0,2} + \beta_{1,2}Age + \beta_{2,2}DAS28 + \beta_{3,2}RF\ +\ \beta_{4,2}sex +\ \beta_{5,2}HAQ$$

Box 1: How to calculate linear predictors and predicted probabilities of a multinomial prediction model (using our multinomial model with three outcomes and five covariates as an illustrative example). *Abbreviations: LP: Linear predictor, LDA: Low disease activity, AEs: Adverse events, DAS28: disease activity score based on 28-joints, RF: Rheumatoid factor, HAQ: Health Assessment Questionnaire*

### *3.2.2. Sample size calculation*

Sample size guidance for developing CPMs for continuous, binary and time-to-event models has been developed in the past few years(50–52), with extensions for the development of a MPM recently proposed(22). Crucially, these sample size calculations help to minimise the level of overfitting of the CPM, and they ensure that there is sufficient sample size to precisely estimate key model parameters (such as the model intercept). A calculation should be performed prior to the development of a MPM, to determine the maximum number of predictor parameters relative to the number of participants, outcome prevalence and expected predictive performance. The total sample size required for a MPM depends on the number of outcome categories, and more outcome categories require larger sample sizes(22). The calculation requires the number of events for each outcome category, the number of candidate predictor parameters, the targeted level of shrinkage, and either the expected concordance statistic (c-statistic) or the expected adjusted $R^2$ for each submodel. Alternatively, if data for model development is collected prospectively, a learning curve approach(53) can be used, potentially in addition to an a priori calculation. See Pate *et al.*(22) for MPM sample size R code and a step-by-step calculation for our clinical example can be found in the Supplementary material.

### *3.2.3. Missing data*

The proportion of missing data in predictor variables should be reported, per variable and overall(54–56). The recommended strategy for handling missing data during model development depends on whether missingness will be present at deployment (i.e. prediction time), as previously proposed by Sisk *et al.(57).* When no missing data is anticipated at model deployment, multiple imputation (including the outcome), or possibly single imputation methods such as regression imputation (omitting the outcome), is recommended for handling missing data during model development(57).

### *3.2.4. Model fitting*

To develop a multinomial model, a MLR is fitted using maximum likelihood estimation (Box 1). The model could also be fit using penalised regression approaches(23), such as LASSO, ridge methods, or Firth's correction(58,59). These may help limit overfitting when coupled with appropriate sample sizes(60). However, the shrinkage parameter itself requires estimation(59,61) so ideally the sample size should be sufficient to not need shrinkage. Continuous predictors should not be categorised during model fitting, to avoid loss of information(62,63).

### *3.2.5. Non-linearity*

Non-linearity in the relationship between continuous predictors and the outcome should be assessed. This is typically done in CPMs using restricted cubic splines (RCS) or fractional polynomials (FP). While some R packages enable the use of RCS with MLR(64,65), current R packages for FPs(66) do not include this functionality for MPMs. In our RA multinomial prediction study, we manually considered fractional polynomial-type powers of continuous variables and compared the Akaike Information Criterion between the different formulations (described in Supplementary material). We did not identify any important non-linear relationships. Alternatively, the FP(66) or RCS(64) function could be applied to each binary submodel but more research, and implementable code, is needed in this area.

### 3.3. Model evaluation

This section will outline performance assessment of MPMs, internal and external validation, and model recalibration.

#### 3.3.1. Performance assessment

##### *3.3.1.1. Calibration*

Calibration quantifies the agreement between predicted risks and observed proportions(67). This is an important performance measure for any CPM as it estimates accuracy of risk predictions. Calibration of an MPM can be assessed following the hierarchy of risk calibration(68).

Level 1: Mean calibration

Mean calibration provides an overall measure of the difference between the observed and expected event rate (calibration-in-the-large). To assess this, the average predicted risk for each outcome is compared with the overall event rate(67), with overestimation occurring when the average predicted risk is higher than the overall outcome prevalence. For MPMs this can be assessed by simply taking the average predicted risk for each outcome category and comparing this to the observed prevalence for that outcome. One can then calculate the Observed/Expected (O/E) ratio; the total number of observed outcome events, divided by the total number of predicted events(51). A ratio <1 suggests an overestimation of risk, a ratio >1 suggests underestimation.

Level 2: Weak calibration

Weak calibration is assessed using the calibration slope (c-slope) and calibration intercept (c-intercept) calculated using the multinomial calibration framework(25), as done in our clinical example. This quantifies whether, on average, the model over- or underestimates risk and does not give risks too extreme or modest. A c-slope <1 indicates that risk estimates are too extreme, and the model may be overfitted, and a c-intercept >0 indicates that the model is underestimating risks(67). C-slopes depend on the choice of reference category and, although this can lead to slightly different results, similar conclusions can be drawn(25). The c-slope is particularly important at model development, as it can be used to shrink model coefficients(69,70) (once adjusted for in-sample optimism). At external validation, the intercept and slope provide a general summary of potential problems with risk calibration(67,68).

In our clinical example, the optimism-adjusted c-slope for the disease activity outcomes (Submodel 1: LDA vs no LDA) was 1.01 [95% confidence interval: 0.87, 1.14] in the development data, which decreased to 0.78 [0.64, 0.93] upon external validation. A decrease in c-slope upon external validation could be linked to differences in underlying patient populations (case-mix) between the development and validation data(68). The c-intercept for Submodel 1 was 0.00 [-0.11, 0.11] in the development data and 0.53 [0.41, 0.65] in the external data, where the difference could be linked to the difference in outcome prevalence between development and validation datasets (more in Section 3.3.3.2.).

The c-slope and c-intercept can also be approximated by obtaining them for each outcome category as if it were binary(12). This simplification yields results that do not depend on reference category.

Level 3: Moderate calibration

A risk model is moderately calibrated if among patients with the same predicted risks, the observed event rate equals the predicted risks. Moderate calibration can reveal miscalibration that is not picked up by weak calibration. This is assessed using calibration plots, which display whether there is any overestimation or underestimation of risk(25). For a MPM, the calibration plots show scatter of the multidimensional relationship between predicted risks and observed proportions, rather than a one-on-one relationship as for a binary outcome(25). For perfect calibration, the scatter should align on the 45-degree line. Calibration plots can be generated using vector spline smoothing(25,71,72). We present the calibration plot at external validation of our case study in Figure 1.

#### *3.3.1.2. Discrimination*

Discrimination is the ability of a CPM to distinguish between patients with different outcome categories. For MPMs, discrimination can be measured using the Polytomous Discrimination Index (PDI), an extension of the c-statistic to obtain simultaneous discrimination between all outcome categories(24). The lower limit of the PDI, which indicates random discrimination, is calculated by $1/k$(24). In our example, the PDI was 0.49 in the development setting (lower limit = 0.33).

For MPMs, c-statistics between pairs of categories can be calculated, with values of 1 and 0.5 indicating perfect and random discrimination, respectively. In our example, Submodel 1 (LDA vs no LDA) had a pairwise c-statistic of 0.72 [0.70, 0.75] in the development setting and 0.68 [0.65, 0.71] in external validation. To obtain pairwise c-statistics that do not involve the reference category (as in the submodels), the easy-to-use conditional risk method is recommended(27).

#### *3.3.1.3. Net benefit*

The clinical utility of a CPM can be assessed using net benefit and decision curve analysis(73,74), and interpreted following the existing step-by-step guide(75). Further research is needed to extend these metrics for MPMs.

### 3.3.2. Internal validation

Following the fitting of the MPM, the model's predictive performance should be quantified through appropriate internal validation techniques to adjust for in-sample optimism. Bootstrapping is the recommended approach as it resamples with replacement, making all data available for model development (contrary to splitting data into development/validation portions). The process for bootstrap internal validation of a MPM is the same as for other types of CPMs; our internal validation procedure for the RA model is reported in the complementary paper(28). One may also consider cross-validation for the internal validation process.

### 3.3.3. External validation

This section describes the external validation of a MPM. The concepts apply to any method, MLR or otherwise, that generates risk estimates for a nominal clinical outcome.

#### *3.3.3.1. Sample size calculation*

Prior to the external validation of a CPM, it is recommended to conduct sample size calculations to obtain the minimum requirements for precise estimates of calibration (O/E, c-slope), discrimination (c-statistic), and clinical utility (net benefit) in the external data(51). Sample size criteria for the external validation of CPMs have been proposed for binary and continuous outcomes(50,51), but not yet extended to MPMs. In our example, we therefore adapted the criteria by calculating the minimum required sample size for each submodel (one-vs-one) and choosing the larger value, which depends on the reference category. These calculations require the end user to specify measures such as outcome prevalence, target standard error (SE) and/or mean and standard deviation of the LP. The expected outcome prevalence and mean/standard deviation of the LP can be informed by the development data, and the target SE relates to the confidence interval width (precision) that one would like to obtain in the external data (worked example in Supplementary material). We recommend that future research considers specific sample size criteria for externally validating a MPM.

#### *3.3.3.2. External validation analysis*

Summaries of baseline characteristics and outcome prevalence(76,77) should be produced to compare case-mix between development and validation settings (as per TRIPOD). It is important to understand heterogeneity between settings as this impacts model performance at external validation. In our clinical example, patients in the validation data had greater disease burden, with a higher disease activity score (median 4.6 [IQR: 3.6-5.4]) compared to the development data (4.2 [3.3-5.2]) and a higher disability index (1.6 [1.1-2.0] vs 1.0 [0.5-1.6]). The outcome prevalence for no LDA, LDA, and discontinuation due to AEs was 40%, 45%, and 15% in the validation data and 45%, 46%, and 9% respectively at development. This shows a particular difference in the prevalence of the AEs outcome, which was ~1.5 times higher in the validation data. Higher heterogeneity between development and validation can impact model calibration, as reported in Section 3.3.1.1.

Missing data should be handled in the same way at validation as is intended when the model is used in practice(56,57). If no missingness is allowed at deployment, but the external validation data contains missing values, this should be imputed using the same approach as for the model development data. Where missingness is allowed at deployment, a consistent imputation model between development, validation, and deployment is recommended(57,78).

For each individual in the external data, the LP and predicted probability for each submodel should be calculated using the equation exactly as developed (Box 1). The same performance metrics as described above for model development can be estimated, and we recommend a particular focus on calibration curves. We present the multinomial calibration plot of our RA model at external validation in Figure 1, which highlights some overprediction of the 'no LDA' category, and underprediction of the 'LDA' category.

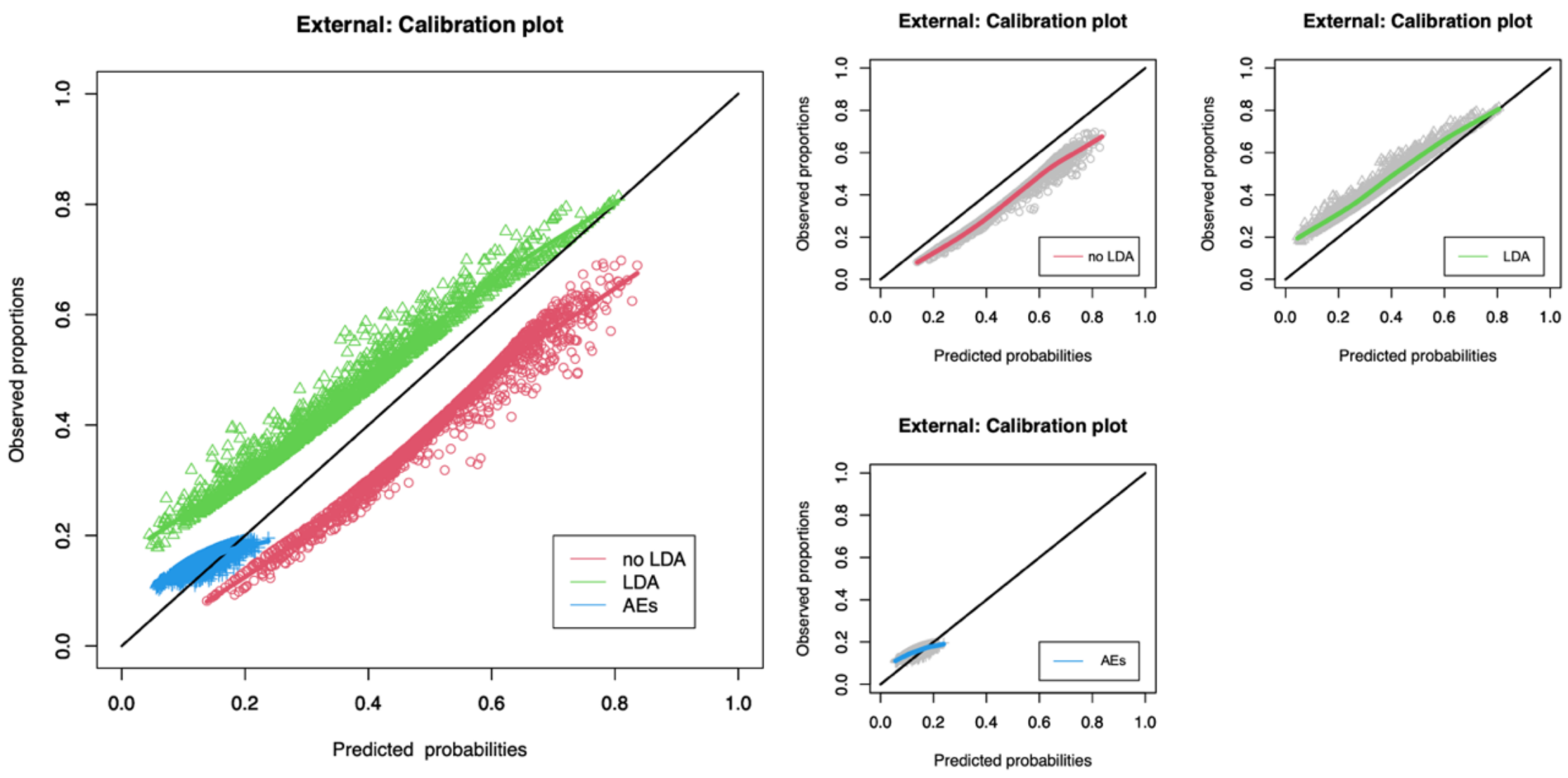

Figure 1: Multinomial calibration plot, using the external performance of our multinomial model for outcomes of methotrexate therapy as an example. *Abbreviations: LDA: Low disease activity, AEs: Adverse events.*

#### 3.3.4. Model recalibration

Upon externally validating a prediction model, it is not uncommon to find a deterioration in performance relative to the development setting. Model updating methods have been proposed as a way of tailoring an existing CPM to suit a new target population(79). These include updating the models intercept and/or slope, refitting the model in the external data, or adding predictors(76). Model updating methods have recently been expanded into the multinomial setting(25,26). In our example, we explored updating the model in two stages(26): 1) recalibration, where the model's slopes and intercepts were updated in the external data, which involves proportionally adjusting the original coefficients (80), and 2) model refitting, where model coefficients were re-estimated in the external data. The approach to recalibration of a MPM is outlined in Box 2, for our RA example.

The LPs of the original multinomial model are denoted by:

$$log\left(\frac{P(Y=2)}{P(Y=1)}\right) = \beta_{0,1} + \beta_{1,1}x_1 + \beta_{2,1}x_2 + \cdots \beta_{p,1}x_p = LP_1$$

$$log\left(\frac{P(Y=3)}{P(Y=1)}\right) = \beta_{0,2} + \beta_{1,2}x_1 + \beta_{2,2}x_2 + \cdots \beta_{p,2}x_p = LP_2$$

Where Y = 2 vs Y = 1 and Y = 3 vs Y = 1 represent the two submodels in the multinomial model (Submodel 1 and 2, respectively). To update this model in the external dataset, the outcome is regressed onto the existing LPs to obtain the following output:

$$log\left(\frac{P(Y=2)}{P(Y=1)}\right) = \alpha_{0,1} + \alpha_{1,1}LP_1 + \alpha_{2,1}LP_2$$

$$log\left(\frac{P(Y=3)}{P(Y=1)}\right) = \alpha_{0,2} + \alpha_{1,2}LP_1 + \alpha_{2,2}LP_2$$

To recalibrate the model in terms of updating the slope and intercept, the alphas in the equation above are used to estimate the new beta coefficients of the recalibrated model. For the Submodel 1 this can be written as:

$$log\left(\frac{P(Y=2)}{P(Y=1)}\right) = \alpha_{0,1} + \alpha_{1,1}(\beta_{0,1} + \beta_{1,1}x_1 + \beta_{2,1}x_2 + \cdots \beta_{p,1}x_p)$$
$$+\alpha_{2,1}(\beta_{0,2} + \beta_{1,2}x_1 + \beta_{2,2}x_2 + \cdots \beta_{p,2}x_p)$$

And similarly, for Submodel 2. Therefore, the updated coefficient for 'age' in Submodel 1 is equal to:

$$(\alpha_{1,1} \times \beta_{1,1}) + (\alpha_{2,1} \times \beta_{1,2}) = \gamma_{1,1}$$

This is repeated for each covariate in each Submodel. The new/updated intercept for Submodel 1 would be:

$$\alpha_{0,1} + (\alpha_{1,1} \times \beta_{0,1}) + (\alpha_{2,1} \times \beta_{0,2}) = \gamma_{0,1}$$

The recalibrated model can be written as:

$$log\left(\frac{P(Y=2)}{P(Y=1)}\right) = \gamma_{0,1} + \gamma_{1,1}x_1 + \gamma_{2,1}x_2 + \cdots \gamma_{p,1}x_p$$

$$log\left(\frac{P(Y=3)}{P(Y=1)}\right) = \gamma_{0,2} + \gamma_{1,2}x_1 + \gamma_{2,2}x_2 + \cdots \gamma_{p,2}x_p$$

Box 2: How to recalibrate (update the slope and intercept) of a multinomial prediction model, using a model with three outcome categories ($Y$).

### 3.4. Computational guidance

Box 3 gives suggestions for R packages(40) for each stage of model development and validation. This is not intended to be an exhaustive list, but lists the packages that we used when developing and evaluating our RA model(28).

- Baseline characteristics: *tableone*(81) for generating baseline tables and quantifying percentage of missing data
- Variable selection: *cv.glmnet()* in *glmnet*(82) for LASSO and ridge; *step()* in *stats(40)* for forward, backward, and stepwise selection;; *logistf* for Firth(83)
- Non-linearity: *fp*(66) for considering FPs for each binary submodel (a form of backward selection, not yet extended for MPMs), *VGAM*(65) for fitting vector splines and *rms(64)* to fix knot locations across the MLR equations
- Sample size calculations: as of December 2023 not yet implemented into an R package; code for the calculation can be found in article by Pate *et al.*(22), which requires the *pROC*(84) library.
- Missing data: *mice*(85) for multiple imputation
- Model fitting: *multinom()* in *nnet*(86), and *VGAM*(65)
- Predictive performance:
    - Calibration: R code for multinomial calibration framework can be found in paper by van Hoorde *et al.*(25), which requires *VGAM*(65) and *bayesm*(87) libraries.
    - Discrimination: Pairwise c-statistics using *pROC*(84) and code for PDI in paper by Dover *et al.(88).* Standard errors can be computed using *mcca (89).*

Box 3: R packages for each stage of model development and performance assessment of a multinomial prediction model

## 4. Conclusion

This article provides an overview and practical guidance on how to develop, externally validate, and update MPMs. Recent methodological advances such as the sample size calculation for the development of a MPM(22), ways to quantify discrimination(24) and calibration of MPMs(25) are enabling the greater use of this approach, but a guide of how to implement many of the ideas and methods is currently lacking. We emphasise that existing best practice guidelines for CPMs should be followed (22,38,39,51,54,76). In this article we outlined MPM-specific considerations only. Future research could focus on MPM considerations for variable selection, fractional polynomials, net benefit, and sample size criteria for external validation.

**Supplementary material**

**How to develop, externally validate, and update multinomial prediction models**

Fractional polynomials sensitivity analysis

In the absence of any R packages that support fractional polynomials (fp) for models based on a multinomial logistic regression, we manually considered powers of continuous variables. For this, we fitted a series of multinomial models for each covariate in turn with varying levels of the power. Using the covariate 'age' as an example, this consisted of modelling:

i) no transformation (just multinom_outcome ~ age)

ii) ii) multinom_outcome ~ age + log(age)

iii) iii) multinom_outcome ~ age + $age^2$

iv) iv) multinom_outcome ~ age + $age^3$

We calculated the AIC of each of the above models. If models with non-linear transformations had a lower AIC to the model without any transformations, this transformation was included in the development of the multinomial model. In our RA example, we did not identify any non-linear relationships.

Step-by-step sample size calculation for the development of a multinomial prediction model for RA, using the calculation outlined by Pate *et al.* (22).

In our RA example, the prevalence of each outcome category was included in the calculation: 1) no LDA = 756 (46%), 2) LDA = 730 (45%), and 3) discontinuation due to AEs = 146 (9%). The shrinkage factor was set to 0.9 and the expected value of the maximum Cox-Snell $R^2$ was set to 15%. Given this information, the total number of predictor parameters that could be considered for the fixed sample size of 1,632 can be established. In our clinical example, this meant that the number of predictors was changed until the total sample size given by the calculation was less than or equal to 1,632. The calculation suggested that eight predictor parameters could be used during model development. Below we outline the final step-by-step calculation using 8 predictor parameters.

**Step 1:** Choose number of candidate predictors (Q). We chose 8.

**Step 2:** Identify sensible values for:

(ii) $p_k$: outcome prevalence (EV) and total number of observations (n)

EV1 = 756, EV2 = 730, EV3 = 146

n = EV1 + EV2 + EV3 = 1632

$p_k$ :

$p_1$ = EV1 / n = 756 / 1632 = 0.46

$p_2$ = EV2 / n = 730 / 1632 = 0.45

$p_3$ = EV3 / n = 146 / 1632 = 0.09

(iii) $p_{k,r}$: outcome prevalence for each submodel (outcome pair)

$p_{1,2}$ = (EV1 + EV2) / n = (756 + 730) / 1632 = 0.91

$p_{1,3}$ = (EV1 + EV3) / n = (756 + 146) / 1632 = 0.55

$p_{2,3}$ = (EV2 + EV3) / n = (730 + 146) / 1632 = 0.54

(iv) max($R^2_{cs_app}$): maximum value of Cox-Snell R2

$\max(R^2_{cs_app}) = 1 - ((p_1\text{^ } p_1)*(p_2\text{^ } p_2)*(p_3\text{^ } p_3))\text{^}2$

= 1 - ((0.46^0.46)*(0.45^0.45)*(0.09^0.09))^2

= 0.85

(v) $R^2_{cs_adj}$: adjusted Cox-Snell R2 (0.15* max($R^2_{cs_app}$))

$R^2_{cs_adj} = 0.15 * \max(R^2_{cs_app})$

= 0.13

(vi) $R^2_{cs_adj,k,r}$ pairwise outcome proportions

phi.1.2 = EV2/(EV1 + EV2) = 730/(756 + 730) = 0.49

phi.1.3 = EV3/(EV1 + EV3) = 146/(756 + 146) = 0.16

phi.2.3 = EV3/(EV2 + EV3) = 146/(730 + 146) = 1.17

$R^2_{cs_adj,1,2}$ = 0.15 * (1 - ((phi.1.2^(phi.1.2))*((1-phi.1.2)^(1-phi.1.2)))^2)

= 0.15 * (1 - ((0.49^(0.49))*((1-0.49)^(1- 0.49)))^2)

= 0.11

$R^2_{cs_adj,1,3}$ = 0.15 * (1 - ((phi.1.3^(phi.1.3))*((1-phi.1.3)^(1-phi.1.3)))^2)

= 0.15 * (1 - ((0.16^(0.16))*((1-0.16)^(1- 0.16)))^2)

= 0.09

$R^2_{cs_adj,2,3}$ = 0.15 * (1 - ((phi.2.3^(phi.2.3))*((1-phi.2.3)^(1-phi.2.3)))^2)

= 0.15 * (1 - ((0.17^(0.17))*((1-0.17)^(1- 0.17)))^2)

= 0.09

**Step 3:** Criterion 1 – targets the global shrinkage factor to be above a pre-specified threshold

(i) Set level of targeted shrinkage (S = 0.9)

(ii) Calculate $m_{k,r}$

$m_{1,2} = Q/((S - 1)*\log(1 - R^2_{cs_adj,1,2} /S))$

$= 8/((0.9 - 1)*\log(1 – 0.11 /0.9))$

$= 599.18$

$m_{1,3} = Q/((S - 1)*\log(1 - R^2_{cs_adj,1,3} /S))$

$= 8/((0.9 - 1)*\log(1 – 0.09 /0.9))$

$= 776.35$

$m_{2,3} = Q/((S - 1)*\log(1 - R^2_{cs_adj,2,3} /S))$

$= 8/((0.9 - 1)*\log(1 – 0.09 /0.9))$

$= 767.54$

(iii) Calculate $n_{kr}$, the total number of individuals needed to achieve the required number of distinct logistic regression model (k,r)

$n_{1,2} = m_{1,2} / p_{1,2} = 599.18 / 0.91 = 658.04$

$n_{1,3} = m_{1,3} / p_{1,3} = 776.35 / 0.55 = 1404.66$

$n_{2,3} = m_{2,3} / p_{2,3} = 767.54 / 0.54 = 1429.94$

(iv) Take the ceiling of the maximum of these as the sample size for criterion 1

**Step 4:** Criterion 2 – targets a small difference (0.05) between the apparent and adjusted Nagelkerke $R^2$.

$NC_2 = 4*Q/(( R^2_{cs_adj} /( R^2_{cs_adj} + 0.05* \max(R^2_{cs_app})) - 1)*\log(1 - R^2_{cs_adj} - 0.05* \max(R^2_{cs_app}))$

$= 4*8/((0.13 /(0.13 + 0.05*0.85) - 1)*\log(1 – 0.13 - 0.05*0.85)$

$= 692$

**Step 5:** Criterion 3 – targets a precise estimate of risk (model intercept) in the overall population

$NC_{3.1}$ = qchisq(1-0.05/5, 1)* $p_1*(1- p_1)/0.05^2 = 659.90$

$NC_{3.2}$ = qchisq(1-0.05/5, 1)* $p_2*(1- p_2)/0.05^2 = 656.12$

$NC_{3.3}$ = qchisq(1-0.05/5, 1)* $p_3$*(1- $p_3$)/0.05^2 = 216.19

Take the ceiling of the maximum number for sample size for criterion 3

**Step 6:** Take the maximum sample size across Criterion 1 – 3

Criterion 1: n = 1430

Criterion 2: n = 691

Criterion 3: n = 660

In our clinical example, criterion 1 was driving the minimum sample size requirement.

External validation sample size calculation

Submodel 1 minimum sample size:

- Observed/Expected (expected outcome prevalence (phi) = 0.45, target standard error (SE) set to 0.051): n=470
- Calibration slope (expected LP mean [sd] = 0.02 [0.9], target SE = 0.072): **n=1518**
- Concordance statistic (expected c-statistic = 0.72, phi = 0.45, target SE = 0.0255): n=400

Submodel 2 minimum sample size results:

- Observed/Expected (expected phi = 0.10, target SE = 0.102 (SE less stringent due to the low phi)): n=865
- Calibration slope (expected LP mean [sd] = -0.33 [0.9], target SE = 0.072): **n=1542**
- Concordance statistic (expected c-statistic = 0.53, phi = 0.1, target SE = 0.03. Less stringent with SE due to low phi): n=1025

Therefore, the minimum sample size (which is the maximum number across all results reported above) is: **n=1542** which is driven by the calibration slope criteria.